\pdfoutput=1
\documentclass[11pt]{article}
\usepackage[T1]{fontenc}
\usepackage[utf8]{inputenc}
\usepackage{lmodern}
\input{glyphtounicode}
\usepackage[margin=1in]{geometry}
\usepackage{graphicx}
\usepackage[hidelinks]{hyperref}
\usepackage{enumitem}
\usepackage{array}
\usepackage{xcolor}
\usepackage{tikz}
\usetikzlibrary{arrows.meta,positioning}
\setlist{noitemsep,topsep=4pt,parsep=0pt,partopsep=0pt}
\newcolumntype{L}[1]{>{\raggedright\arraybackslash}p{#1}}
\definecolor{mazoBlue}{HTML}{EAF2FF}
\definecolor{mazoGreen}{HTML}{EAF7EA}
\definecolor{mazoGold}{HTML}{FFF5D6}
\definecolor{mazoRed}{HTML}{FDECEC}
\definecolor{mazoGray}{HTML}{F4F4F4}
\hypersetup{
  pdftitle={Mazocarta: A Seeded Procedural Deckbuilder for Instrumented Game Development},
  pdfauthor={Timothy C. Cogan}
}

\title{Mazocarta: A Seeded Procedural Deckbuilder \\ for Instrumented Game Development}
\author{Timothy C. Cogan \\ Independent Researcher \\ tim@cogan.dev}
\date{}

\begin{document}
\maketitle

\begin{abstract}
Mazocarta is a seeded procedural tactical deckbuilder implemented in Rust, compiled to WebAssembly for
browser play, and executable natively for simulation.
Its primary technical contribution is not the invention of a new deckbuilding genre, but the construction of an
instrumented game-development reference artifact: the same rules engine supports interactive play, native
command-line simulation, automated end-to-end tests, save/load fixtures, and local-area multiplayer.
This paper describes Mazocarta's architecture, deterministic run model, reproducible balance probes, and
QR-mediated WebRTC pairing for local multiplayer.
An evaluation snapshot over 1,000 deterministic seeds shows that the simulation pipeline can produce reproducible
development signals. In the evaluated configuration, single-player and two-player autoplay win rates were 36.1\%
and 34.9\% over 1,000 deterministic seeds, respectively. These rates are not presented as final player-facing
balance metrics, but as repeatable probes for future balance shifts and regressions.
Mazocarta is positioned as a playable open-source reference artifact for instrumented game development:
deterministic regression checks, automated playtesting workflows, balance probes for game mechanics, and
browser-native local multiplayer all exercise one shared production rules core.
\end{abstract}

\section{Introduction}
Run-based procedural deckbuilders combine seeded map progression, hidden and visible random events, card synergies,
enemy behaviors, and long-run economy decisions. These numerous, complex features combine into emergent game-experiences
which make mechanics difficult to balance and validate by
manual play alone: small rule or content changes can affect combat outcomes, map progression, reward selection,
and late-run difficulty in ways that may not be visible in a short human test session.

Mazocarta addresses this problem as a small but complete game architecture rather than as a standalone
balancing algorithm. The game is implemented in Rust~\cite{rust-lang}, compiled to WebAssembly for browser
play~\cite{webassembly,haas2017bringing}, and exposed through a thin JavaScript host. The same Rust rules can also be executed
natively by a command-line actor, allowing deterministic runs over many seeds without browser rendering.
This creates a direct path from interactive implementation to automated playtesting and balance probing: the
artifact being measured is the artifact being played.

This paper presents Mazocarta as an open-source software system~\cite{mazocarta-software} for instrumented game
development over deterministic procedural rules. It follows prior work that uses autonomous agents and procedural
personas for automated playtesting~\cite{yannakakis2018artificial,holmgaard2018automated,zook2019automatic},
but focuses on a complementary engineering problem. In many playtesting settings, researchers do not control the
full production codebase, or the game was not designed to expose its rules as a reusable API. Playtesting tools may
therefore rely on wrappers, telemetry, UI-level control, abstract models, or separate approximations of gameplay
logic. Mazocarta instead asks how an independent game can be structured so that balance probes, regression tests, UI
tests, and multiplayer synchronization all exercise the same production rules core.

\paragraph{Contributions}
This paper makes three practical contributions:
\begin{itemize}
  \item A seeded deterministic Rust rules core with an end-to-end verification surface spanning unit tests,
  save/load fixtures, Playwright browser tests, fake-camera QR tests, and multiplayer soak runs.
  \item A lightweight autoplay and native actor pipeline for reproducible game-balance and progression probes over
  fixed seed ranges.
  \item A QR-mediated WebRTC local multiplayer model that separates shared dungeon state from per-hero state without
  requiring a hosted signaling or game server.
\end{itemize}

\section{Motivation and Design Goals}
Mazocarta is shaped by three design goals.
\begin{itemize}
  \item \textbf{Deterministic rules.} Combat, deck order, map generation, rewards, events, and enemy behavior should be
  reproducible from explicit seeds and serialized state.
  \item \textbf{Playable instrumentation.} Test and simulation hooks should exercise the same rules used by players, rather
  than a simplified analytical model maintained separately from the game.
  \item \textbf{Local-first deployment.} The game should run in a browser, support offline installation after the first load,
  and allow local multiplayer without requiring a hosted game server~\cite{kleppmann2019local}.
\end{itemize}

These goals lead to a conservative architecture. Mazocarta does not attempt to provide a general game engine or
a general automated balancing framework. Instead, it packages a specific deckbuilder with enough deterministic
surface area to support repeatable playtesting, debugging, and regression checks.

\section{System Overview}
Mazocarta is a tactical card game with an $N$-sector run structure, branching map progression, visible enemy
intents, status axes, modules, consumables, shops, events, rest nodes, elite fights, and boss encounters.
The sector count is a tuning parameter rather than a fixed rule assumption.
The interface currently ships with English and Spanish text, with the same content tables adaptable to additional
languages, and the browser client supports mouse, keyboard, and touch input.

The combat model centers on three signed axes: Focus scales outgoing attack damage, Rhythm scales shield gain,
and Momentum scales energy and enemy intent output. These axes do not form a hidden global rock-paper-scissors
table. Instead, Mazocarta uses them as a lightweight interaction system: cards, enemy intents, rewards,
requirements, and modules read from and write to the same signed values, scale ordinary effects from them, or
replace those effects altogether. Because these relationships are content-defined rather than hard-coded, downstream
developers can retune the current pressure pattern or replace card, enemy, reward, and module effects for their own
balance goals without changing the deterministic rules surface. The result is a compact rules vocabulary that can
be displayed to players while still creating combinatorial interactions across deck composition, enemy intent
cycles, and module choices.

\begin{figure}[t]
\centering
\includegraphics[width=0.96\linewidth]{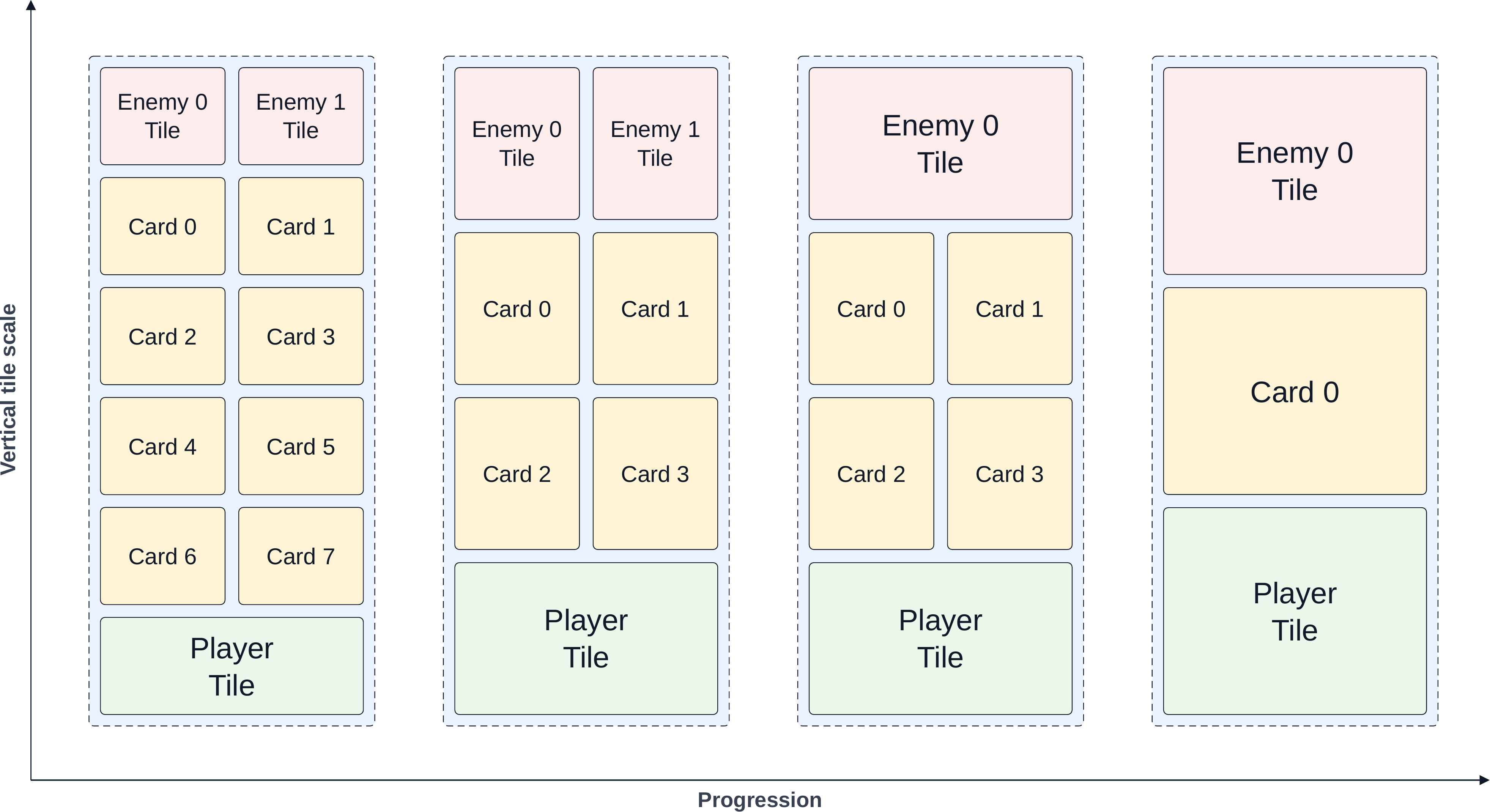}
\caption{Responsive tile-based combat surface. On narrow screens, Mazocarta keeps combat state in stable tiles while
reallocating free space as encounters progress: larger hands use compact cards, and smaller hands expand card tiles
as enemies leave play to preserve readability and touch targets.}
\label{fig:combat-surface}
\end{figure}

The combat presentation uses a compact tile-based visual language: cards, enemy panels, status readouts, and action
targets are represented as stable geometric units rather than as asset-heavy animation. These units reflow and scale
across viewport and hand-size changes, keeping player-facing state readable while making interaction targets and
end-to-end test fixtures straightforward to expose and validate. Figure~\ref{fig:combat-surface} shows the intended
abstraction: the interface is not merely decorative, but a structured view of state that adapts to available space.

\begin{figure}[t]
\centering
\includegraphics[width=0.98\linewidth]{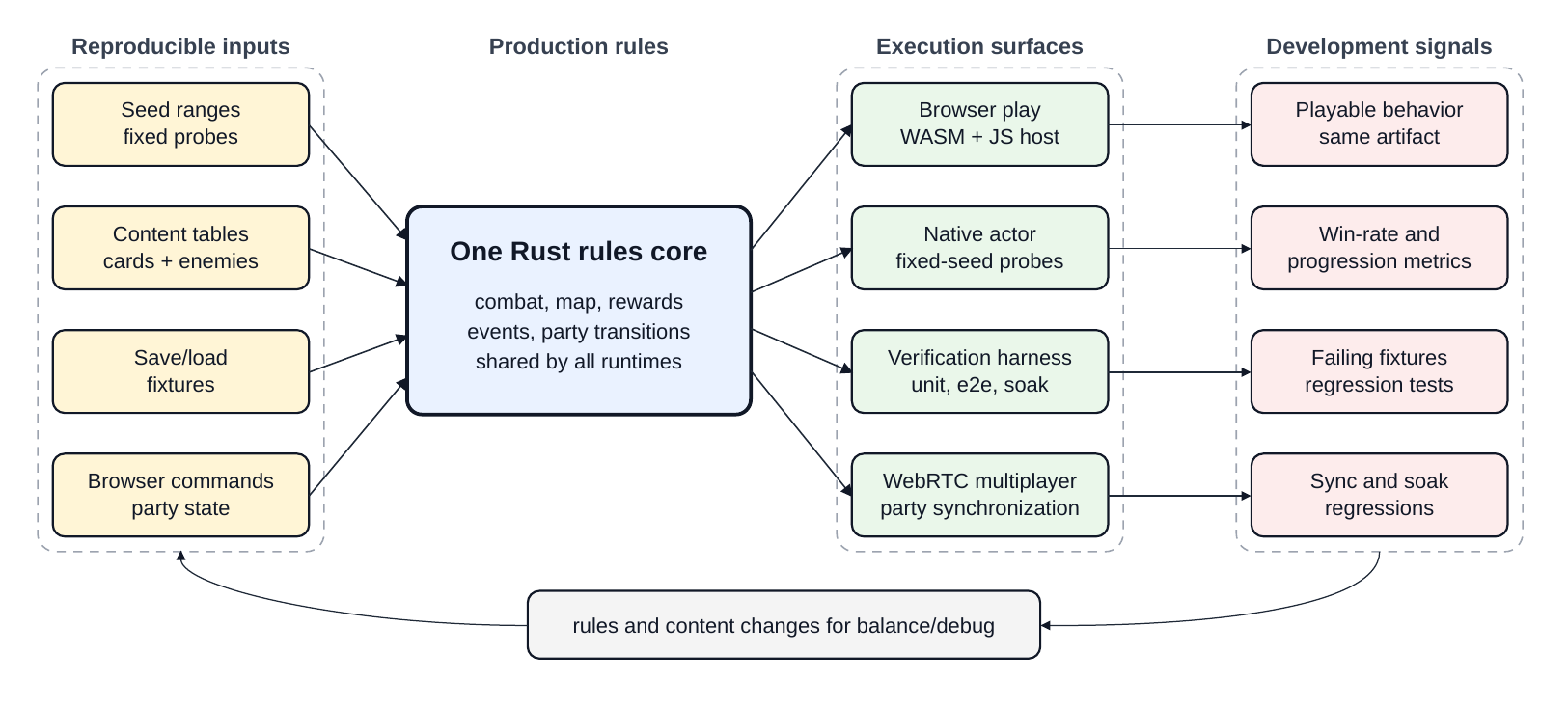}
\caption{Instrumentation loop. Reproducible inputs enter one production Rust rules core; the same
state transitions drive browser play, native actor probes, verification harnesses, and multiplayer sessions, which
return concrete development signals for balance, debugging, and regression detection.}
\label{fig:instrumentation}
\end{figure}

Figure~\ref{fig:instrumentation} summarizes the implementation as an instrumentation loop. Reproducible seeds,
content tables, fixtures, and party state enter one production Rust rules core. That core drives browser play,
native actor probes, verification harnesses, and WebRTC multiplayer sessions; those surfaces then return concrete
development signals for playability, balance, debugging, and regression detection.

\section{Deterministic Rules and Simulation}
Mazocarta's simulation value comes from sharing state transitions across interactive and automated contexts.
The actor simulator starts a run at a fixed seed, chooses map nodes, resolves rewards, handles shops and events,
and plays combat through a heuristic policy. The policy is intentionally simple: it scores legal card actions,
prioritizes victory, defensive survival, useful zero-cost play, setup actions, and positive trades, then advances
the same combat state that the browser client would advance.

This differs from a learned player model or Monte Carlo tree search persona~\cite{yannakakis2018artificial,holmgaard2018automated}. The current
actor is best understood as a deterministic heuristic baseline and probe: it is not a claim about optimal play or
human play, but a stable instrument for detecting gross difficulty shifts, unreachable states, freezes, and
regressions in reward or map logic. More capable reinforcement-learning, search-based, or persona-based agents
should outperform it as players; its value here is that it is cheap, reproducible, and directly connected to the
production rules. Because the actor executes those rules, a failed or stalled run usually points to a concrete
implementation or balance issue rather than to mismatch between a model and the game.

The same philosophy appears in the browser tests. Playwright-based end-to-end tests load deterministic fixtures,
exercise combat input, validate host/guest pairing, scan animated QR frames through a fake camera path, and verify
that a multiplayer guest can observe the host's map state. A separate soak runner repeatedly pairs two browser pages,
starts deterministic two-player runs, autoplays both clients, and reports the first seed that stalls.

\section{Local-First Multiplayer}
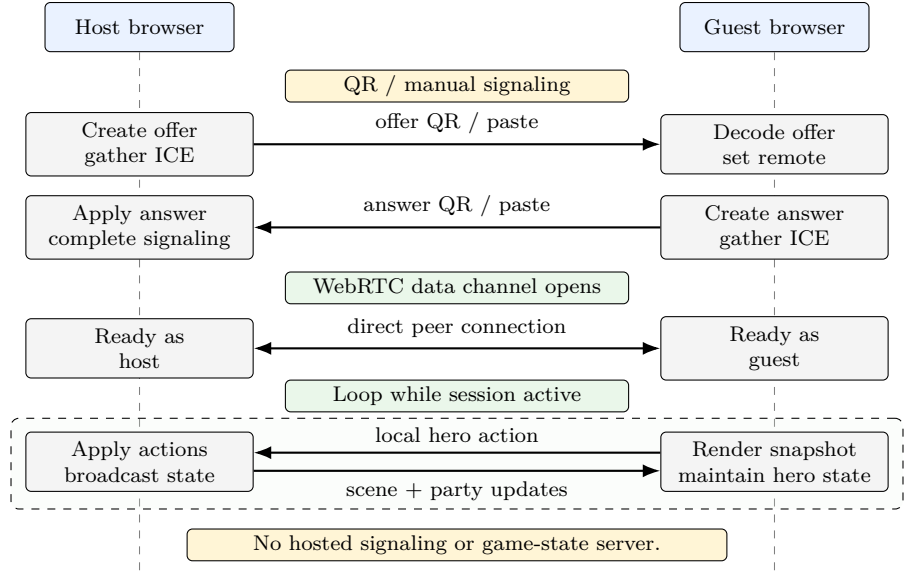
\begin{figure}[t]
\centering
\begin{tikzpicture}[
  font=\scriptsize,
  role/.style={draw, rounded corners=2pt, fill=mazoBlue, align=center, minimum width=2.5cm, minimum height=0.65cm},
  seqstep/.style={draw, rounded corners=2pt, fill=mazoGray, align=center, text width=2.75cm, minimum height=0.72cm},
  phase/.style={draw, rounded corners=2pt, fill=mazoGold, align=center, text width=4.4cm, minimum height=0.42cm, inner sep=2pt},
  channel/.style={draw, rounded corners=2pt, fill=mazoGreen, align=center, text width=4.4cm, minimum height=0.42cm, inner sep=2pt},
  loopbox/.style={draw, dashed, rounded corners=3pt, fill=mazoGreen, fill opacity=0.18, inner sep=0pt},
  msg/.style={->, thick, >=Latex},
  lifeline/.style={gray, dashed}
]
\node[role] (host) at (0,0) {Host browser};
\node[role] (guest) at (8.4,0) {Guest browser};
\draw[lifeline] (host.south) -- ++(0,-6.9);
\draw[lifeline] (guest.south) -- ++(0,-6.9);

\node[phase] (signal) at (4.2,-0.78) {QR / manual signaling};
\node[seqstep] (hOffer) at (0,-1.55) {Create offer\\gather ICE};
\node[seqstep] (gOffer) at (8.4,-1.55) {Decode offer\\set remote};
\draw[msg] (hOffer.east) -- node[above, align=center] {offer QR / paste} (gOffer.west);

\node[seqstep] (hAnswer) at (0,-2.65) {Apply answer\\complete signaling};
\node[seqstep] (gAnswer) at (8.4,-2.65) {Create answer\\gather ICE};
\draw[msg] (gAnswer.west) -- node[above, align=center] {answer QR / paste} (hAnswer.east);

\node[channel] (data) at (4.2,-3.45) {WebRTC data channel opens};
\node[seqstep] (hOpen) at (0,-4.25) {Ready as\\host};
\node[seqstep] (gOpen) at (8.4,-4.25) {Ready as\\guest};
\draw[msg,<->] (hOpen.east) -- node[above, align=center] {direct peer connection} (gOpen.west);

\node[channel] (loopLabel) at (4.2,-4.88) {Loop while session active};
\node[loopbox, minimum width=11.8cm, minimum height=1.20cm] (sessionLoop) at (4.2,-5.775) {};
\node[seqstep] (hState) at (0,-5.75) {Apply actions\\broadcast state};
\node[seqstep] (gState) at (8.4,-5.75) {Render snapshot\\maintain hero state};
\draw[msg] ([yshift=0.12cm]gState.west) -- node[above, align=center] {local hero action} ([yshift=0.12cm]hState.east);
\draw[msg] ([yshift=-0.12cm]hState.east) -- node[below, align=center] {scene + party updates} ([yshift=-0.12cm]gState.west);
\node[phase, text width=7.0cm] at (4.2,-6.85) {No hosted signaling or game-state server.};
\end{tikzpicture}
\caption{QR-mediated WebRTC pairing. Mazocarta uses QR frames or manual paste only for signaling; after the
offer/answer exchange, the peer connection carries game-state updates directly between browsers.}
\label{fig:qr-flow}
\end{figure}

Mazocarta's multiplayer path is designed for local-area cooperative play. WebRTC peer connections provide the
browser-to-browser transport; as with standard WebRTC, peers exchange offer/answer descriptions and ICE information
through a signaling path before the peer connection can carry data~\cite{webrtc-w3c,keranen2018interactive,webrtc-peer-connections,mdn-webrtc-connectivity}.
Mazocarta uses QR codes and manual paste as that signaling path instead of a central matchmaking server.
Figure~\ref{fig:qr-flow} summarizes this connection flow.

The pairing bridge serializes offer and answer payloads, optionally compresses them, encodes them as QR-friendly text,
and splits large payloads into animated transport frames. This supports phones and laptops on the same local network:
the host displays an offer, the guest scans or pastes it, the guest returns an answer, and the host applies it.
After connection, the host broadcasts scene and party state while each participant keeps a per-hero run state.

The multiplayer model separates shared dungeon state from per-hero state. The party shares map structure, current
node, visited path, level progression, and room counters; each hero maintains deck, modules, hit points, maximum hit
points, and credits. This allows one map decision to define the party route while combat and reward consequences
remain attributable to individual heroes.

\section{Artifact and Reproducibility}
The evaluated artifact is the public Mazocarta repository~\cite{mazocarta-software}, licensed under MIT, at commit
\texttt{65818ed3ba84942efbbe950099def180f7e3c9ac}. The local reproduction environment used Rust
\texttt{1.93.1}, Node.js \texttt{24.6.0}, npm \texttt{11.12.0}, and Playwright \texttt{1.59.1}.

Table~\ref{tab:evaluation} reports commands that can be run from the repository root. In actor reports, a win is a
completed run, a loss is an ordinary game defeat, and an abort is a simulator guardrail such as missing progression
state, an invalid setup, or exceeding step limits. Combat, elite, and boss averages are over all runs; victory HP and
surviving-hero averages are over wins only. In the browser soak runner, a stall means both pages stop producing
actions for the idle limit. A progress timeout means the run budget elapsed while the session was still advancing; it
is useful for smoke testing synchronization, but it is not evidence of a completed run.

\section{Evaluation Snapshot}
Table~\ref{tab:evaluation} reports a compact reproducibility snapshot from the current implementation. The purpose
is not to establish final game balance, but to show that Mazocarta can produce repeatable aggregate signals from
the same code used in play.
All results were generated from a clean checkout of the evaluated commit.

\begin{table}[t]
\centering
\footnotesize
\begin{tabular}{|L{0.24\linewidth}|L{0.41\linewidth}|L{0.23\linewidth}|}
\hline
\textbf{Check} & \textbf{Command / setup} & \textbf{Observed result} \\
\hline
Rust rule tests & \texttt{cargo test -q} & 466 passed, 0 failed \\
\hline
Save/load tests & \texttt{cargo test -q save} & 25 passed, 0 failed \\
\hline
Browser E2E and QR tests & \texttt{make test-e2e} & 23 passed, 0 failed \\
\hline
Multiplayer soak smoke & \texttt{npm run soak:2p -{}- -{}-runs 5 -{}-seed-start 1} & 0 stalls; 5 progress timeouts while still advancing \\
\hline
Single-player actor & \texttt{cargo run -{}-bin actor -{}- -{}-runs 1000 -{}-seed-start 1} & 361 wins, 639 losses, 0 aborts \\
\hline
Two-player actor & \texttt{cargo run -{}-bin actor -{}- -{}-players 2 -{}-runs 1000 -{}-seed-start 1} & 349 wins, 651 losses, 0 aborts \\
\hline
\end{tabular}
\caption{Evaluation snapshot over deterministic tests and actor simulations. The actor results are treated as
development signals rather than final player-facing balance metrics.}
\label{tab:evaluation}
\end{table}

\begin{figure}[t]
\centering
\begin{tikzpicture}[
  x=0.23cm,
  y=0.82cm,
  font=\scriptsize,
  axis/.style={black!70},
  grid/.style={black!12},
  bar/.style={draw=black!65, rounded corners=1pt}
]
\draw[axis,->] (0,0) -- (42.5,0) node[right] {win rate};
\foreach \x in {0,10,20,30,40} {
  \draw[grid] (\x,0) -- (\x,2.25);
  \draw[axis] (\x,0) -- (\x,-0.08) node[below] {\x\%};
}
\node[anchor=east] at (-0.8,1.65) {Single-player};
\filldraw[bar, fill=mazoBlue] (0,1.4) rectangle (36.1,1.9);
\node[anchor=west] at (36.7,1.65) {36.1\%};

\node[anchor=east] at (-0.8,0.65) {Two-player};
\filldraw[bar, fill=mazoRed] (0,0.4) rectangle (34.9,0.9);
\node[anchor=west] at (35.5,0.65) {34.9\%};
\end{tikzpicture}
\caption{Autoplay win-rate signal over 1000 deterministic seeds. The chart is intended as a development probe,
not as evidence of final player-facing balance.}
\label{fig:autoplay-bars}
\end{figure}
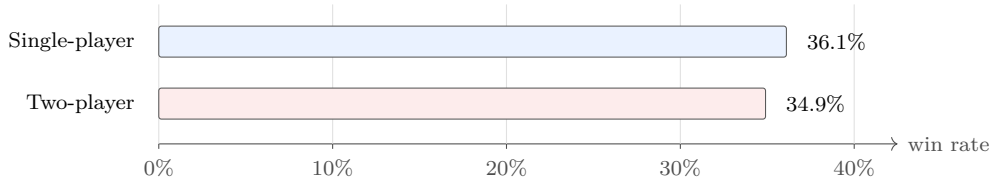

Figure~\ref{fig:autoplay-bars} visualizes the same result as a balance signal. The actor is deliberately simple,
but the fixed-seed pipeline supports reproducible comparison across implementation changes without changing the
playable implementation.

The single-player actor completed 1000 runs with no simulation aborts, clearing on average 14.85 combats, 1.65
elites, and 2.15 bosses, with average victory HP of 37.85. The two-player actor also completed without aborts,
won 349 of 1000 runs, cleared on average 14.03 combats, 1.49 elites, and 2.03 bosses, and finished victories with
average party HP of 78.98 and 2.00 surviving heroes. The one-player and two-player win rates are in the same broad
range and should not be read as evidence that one mode is intrinsically harder. This snapshot shows that the same
instrumentation that supports confidence in rule execution also supports reproducible comparison after
implementation changes.

\section{Related Work}
Mazocarta is most closely related to work on automated playtesting and procedural content evaluation. Procedural
personas use artificial play styles to test game content and expose how different agents experience levels
or mechanics~\cite{holmgaard2018automated}. Active-learning approaches have also been used to reduce the number of
playtests required for parameter tuning~\cite{zook2019automatic}. Mazocarta is less ambitious algorithmically:
it uses hand-written heuristics rather than learned or evolved agents. Its contribution is the integration of such
probes into the production rule implementation of a complete browser game.

Industry-oriented work on autoplay agents for prelaunch game tuning has also argued for instrumented game clients,
agent-facing APIs, and simulations that preserve the relationship between normal game execution and automated
testing~\cite{borovikov2021autoplay}. Mazocarta follows this engineering direction rather than claiming it as new.
Its focus is narrower and artifact-centered: it packages a complete open-source browser game whose interactive
client, native simulator, save/load fixtures, browser tests, and local multiplayer all share one Rust rules core.

The map and reward structure fits within the broader tradition of procedural content generation in games, where
generated spaces and sequences are evaluated according to playability, challenge, and design intent~\cite{togelius2011search,shaker2016procedural}.
Mazocarta's generated maps are deliberately constrained: each run has a configurable finite sector structure,
explicit room types, and deterministic seeds, making simulation traces and regression reproduction straightforward.

On the networking side, Mazocarta uses standard WebRTC peer connection concepts but replaces server-hosted signaling
with QR/manual exchange for local sessions. This does not remove WebRTC's connectivity constraints; NAT behavior,
browser support, and ICE gathering still matter. It does, however, align the deployment model with a local-first
game that can be installed and played without operating a central game service.
Manual serverless WebRTC offer/answer exchange and QR-code variants also predate Mazocarta~\cite{ball2013serverless,serverless-webrtc-qrcode}.
Mazocarta therefore does not claim QR signaling itself as novel. Its contribution is the integration of QR/manual
signaling, deterministic shared game state, per-hero state separation, and automated multiplayer tests inside one
playable deckbuilder artifact.

\section{Discussion}
Mazocarta demonstrates a practical pattern for small deterministic games: implement the rules once, expose them to
both the player-facing runtime and the simulation runtime, and treat automation as a first-class product surface.
This pattern is especially useful for deckbuilders because their interesting behavior emerges over many decisions,
not only within a single combat.

The current implementation also shows the limits of lightweight automation. A stable heuristic actor can detect
major regressions and produce comparable numbers over seed ranges, but it cannot replace human playtesting. A low
autoplay win rate may indicate overtuned content, a weak policy, poor cooperative coordination, or some combination
of these. For this reason, Mazocarta's metrics should be interpreted as reproducible signals for development rather
than as direct measurements of player experience.

The same properties may make Mazocarta useful as a teaching artifact, although that use is not evaluated here.
Because the game is open-source, compact, deterministic, and playable, it can expose concrete examples of procedural
generation, game AI baselines, automated testing, WebAssembly deployment, and local browser networking without
requiring students to work inside a large commercial game engine. Fixed-seed runs, native actor reports, visible
combat state, and end-to-end fixtures give instructors and students inspectable entry points into the system.

\section{Limitations}
The current evaluation is intentionally modest. It reports deterministic tests and aggregate actor outcomes, but
does not include a user study, comparative genre evaluation, ablation study, or statistical analysis across multiple
actor policies. The autoplay policy is hand-written and may underrepresent human planning, especially in two-player
coordination.

The multiplayer design is server-free for live game coordination and signaling: Mazocarta avoids a central game
server and uses QR/manual exchange instead of server-hosted signaling. It still requires a viable browser-to-browser
connection, and WebRTC connectivity can depend on network topology, browser implementation, and ICE behavior. QR
exchange also adds practical usability constraints: camera access, animated frame decode reliability, and payload size
all affect pairing quality.

Finally, because Mazocarta is an evolving game implementation, the reported balance numbers should be interpreted
as a snapshot of the evaluated artifact rather than as stable properties of the game. Future content changes or
actor-policy changes may change the reported win rates, but the fixed-seed instrumentation makes those shifts
reproducible and comparable.

\section{Future Work}
The most important next step is a stronger evaluation harness. Mazocarta should report actor runtime, seed ranges,
confidence intervals, failure classifications, and multiple actor policies with different risk profiles. Two-player
balance still needs separate analysis because comparable aggregate win rates can hide coordination, survivability,
and per-hero contribution problems.

Additional future work includes browser performance benchmarks, offline installation checks, real-device QR pairing
tests, and human playtesting that can compare player outcomes with actor-derived expectations.
Future agent work should treat the current actor as a baseline to be outperformed by reinforcement-learning,
search-based, or persona-based policies while retaining fixed-seed comparability against the same production rules.

\section{Conclusion}
This paper has presented Mazocarta as a seeded procedural deckbuilder for instrumented game development.
Its contribution lies in the integration of an end-to-end verification surface, native actor simulation, and
QR-mediated local multiplayer around one shared Rust rules core. The current evaluation shows how deterministic
automation can exercise rule stability across many runs while also making balance questions easier to reproduce and
inspect. Mazocarta therefore provides a compact open-source example of how game mechanics can be built for play,
testing, balancing, and reproducible analysis at the same time.

\begingroup
\small
\raggedright
\bibliographystyle{unsrt}
\bibliography{refs}
\endgroup

\end{document}